\documentclass[12pt]{article}
\usepackage[reqno]{amsmath}
\usepackage{bbm}
\usepackage{epsfig}
\usepackage{array}
\usepackage{float}

\usepackage{a4}

\usepackage{a4wide}
\usepackage{wasysym}

\def\ra{\rightarrow}
\def\be{\begin{equation}}
\def\ee{\end{equation}}
\def\gs{\mathrel{
   \rlap{\raise 0.511ex \hbox{$>$}}{\lower 0.511ex \hbox{$\sim$}}}}
\def\ls{\mathrel{
   \rlap{\raise 0.511ex \hbox{$<$}}{\lower 0.511ex \hbox{$\sim$}}}}

\newcommand{\ba}{\begin{array}{c}}
\newcommand{\baz}{\begin{array}{cc}}
\newcommand{\bad}{\begin{array}{ccc}}
\newcommand{\bea}{\begin{equation} \begin{array}{c}}
\newcommand{\eea}{ \end{array} \end{equation}}
\newcommand{\ea}{\end{array}}
\newcommand{\D}{\displaystyle}
\newcommand{\dms}{\mbox{$\Delta m^2_{\odot}$}}
\newcommand{\dma}{\mbox{$\Delta m^2_{\rm A}$}}
\newcommand{\meff}{\mbox{$\langle m \rangle$}}

\begin{document}

\title{
\vspace{-2cm}
\hfill {\small SISSA 22/2004/EP}\\
\vspace{-0.3cm}
\hfill {\small hep-ph/0403236} \\ 
\vskip 0.8cm
\bf Type II See--Saw Mechanism, Deviations from Bimaximal Neutrino 
Mixing and Leptogenesis
}
\author{
Werner Rodejohann\thanks{email: \tt werner@sissa.it}\\ \\
{\normalsize \it Scuola Internazionale Superiore di Studi Avanzati}\\
{\normalsize \it Via Beirut 2--4, I-34014 Trieste, Italy}\\
{\normalsize \it and}\\
{\normalsize \it Istituto Nazionale di Fisica Nucleare}\\
{\normalsize \it Sezione di Trieste, I-34014 Trieste, Italy}
}
\date{}
\maketitle
\thispagestyle{empty}
\vspace{-0.8cm}
\begin{abstract}
\noindent 
A possible interplay of both terms in the type II 
see--saw formula is illustrated by presenting 
a novel way to generate deviations from exact bimaximal neutrino 
mixing.   
In type II see--saw mechanism with dominance of the non--canonical 
$SU(2)_L$ triplet term, the 
conventional see--saw term can naturally give a small 
contribution to the neutrino mass matrix. If the 
triplet term corresponds to the bimaximal mixing scheme 
in the normal hierarchy, the small contribution of the conventional see--saw 
term naturally generates non--maximal solar neutrino mixing. 
Atmospheric neutrino mixing is also reduced from maximal, 
corresponding to $1 - \sin^2 2 \theta_{23}$ of order 0.01.
Also, small but non--vanishing $U_{e3}$ of order 0.001 is obtained. 
It is also possible that the $\Delta m^2$ responsible for 
solar neutrino oscillations is induced by the small conventional see--saw 
term. Larger deviations from zero $U_{e3}$ and from maximal atmospheric 
neutrino mixing are then expected. This scenario links the small ratio 
of the solar and atmospheric $\Delta m^2$ with the deviation from 
maximal solar neutrino mixing. 
We comment on leptogenesis in this scenario and compare the 
contributions to the decay asymmetry of the heavy Majorana neutrinos 
as induced by themselves and by the triplet.

\end{abstract}

\newpage

\section{\label{sec:intro}Introduction}
The bimaximal neutrino mixing scheme \cite{bimax} is 
esthetically and theoretically very 
appealing but suffers --- as many simple frameworks --- 
from the fact that Nature did not realize it with perfect 
accuracy. In this particular case, 
solar neutrino mixing is not maximal. 
Recent data as collected by the SNO salt phase \cite{SNO2} 
and by the KamLAND \cite{KL} experiment  
shows that $\tan^2 \theta_{\rm sol} < 1$ at more than 5$\sigma$ \cite{SNO2}. 
Nevertheless, the (close to) maximal atmospheric neutrino mixing and 
the smallness of $U_{e3}$ can lead one to believe 
that Nature presumably started 
with bimaximal mixing and that the underlying symmetry 
(one could imagine, e.g., flavor symmetries such as 
$L_e - L_\mu - L_\tau$ \cite{lelmlt}) was somehow 
broken to yield the observed phenomenology. 
There are several approaches trying to explain deviations from 
bimaximal neutrino mixing, e.g., from charged lepton mixing 
\cite{devbima1,steve,WR,FPR}, from radiative corrections \cite{devbima2} 
or\footnote{It is even possible 
to construct situations in which maximal solar neutrino mixing would imply 
a vanishing baryon asymmetry of the universe \cite{ichEJP}.} 
from breaking of symmetries \cite{flav}. 

\noindent The primary goal of all those approaches 
is to generate non--maximal solar neutrino mixing, which can be 
parametrized through a small parameter $\lambda \sim 0.2$  as \cite{WR} 
\be \label{eq:lam}
U_{e2} = \sqrt{\frac{1}{2}}~(1 - \lambda)~.
\ee
Note that the parameter $\lambda$ is very similar to the Cabibbo angle: 
$\lambda \simeq  \theta_C$ \cite{WR}. 
When one starts from bimaximal neutrino mixing and introduces some 
deviation to generate non--maximal solar neutrino mixing, usually also 
$U_{e3}$ and atmospheric mixing will deviate from their ``bimaximal'' 
values. Thus, those additional deviations will 
be proportional to some power of $\lambda$, depending on the model. 
A useful parametrization taking into account this fact is \cite{WR} 
\be \label{eq:devs}
U_{e3} = A_\nu~\lambda^n \mbox{ and } 
U_{\mu 3} = \sqrt{\frac{1}{2}}~(1 - B_\nu ~\lambda^m)~e^{i \delta}, 
\ee
where the integer numbers $m, n$ and the order one numbers $A_\nu, B_\nu$ 
are to 
be chosen according to the numerical values of $U_{e3}$ and $U_{\mu 3}$.
 
\noindent Numerically, it turns out that the phenomenologically 
very interesting ratio $R$ of the 
solar and atmospheric $\Delta m^2$ is of order $\lambda^2$ \cite{WR}. 
This can be interpreted in the sense that both 
the deviation from bimaximal neutrino mixing {\it and} the small ratio 
$R$ have the same origin. 
Hence, it would be interesting to have a scenario in which one 
starts from zero \dms{} together with bimaximal mixing and ends up with the 
observed phenomenology of non--maximal solar neutrino mixing and a small 
ratio of the solar and atmospheric $\Delta m^2$.

\noindent 
In this note we shall try to reason that in type II see--saw models, 
i.e., 
\be 
m_\nu = m_L - m_D^T \, M_R^{-1} \, m_D  ~,
\ee
with dominance of the $SU(2)_L$ triplet term \cite{jose1} $m_L$, 
the conventional see--saw term \newline $m_D^T \, M_R^{-1} \, m_D$ can 
naturally give a 
small correction to $m_L$. Starting with a structure of $m_L$ 
which corresponds to bimaximal mixing in the normal mass hierarchy  
and using the natural assumptions $m_L \propto M_R$ as well as 
hierarchical $m_D$, one can easily obtain non--maximal solar neutrino 
mixing while keeping the atmospheric neutrino mixing close to maximal 
and generating small but non--vanishing $U_{e3}$. 
In addition, it is also possible to generate non--vanishing \dms{} 
via the small conventional see--saw contribution. 
This is then a scenario linking the deviations from bimaximal 
mixing and the small $R$. 

\noindent 
We shall not speculate about particular models in which such a scenario  
arises but rather investigate its consequences on neutrino phenomenology. 
The simple situation from which we start, the illustration of 
an interesting interplay of both terms in the type II see--saw formula 
and the possibility to link the small ratio $R$ with the deviation from 
bimaximal neutrino mixing justifies this approach.

\noindent We also consider leptogenesis as realized 
through heavy Majorana neutrino decays  
and compare for the first time in a concrete situation 
the contributions to the decay asymmetry induced by diagrams 
involving virtual Majorana neutrinos and the $SU(2)_L$ triplet.\\ 

\noindent The paper is organized as follows. In Section \ref{sec:frame} 
we briefly describe the phenomenological and theoretical framework 
in which we work. We consider the $CP$ conserving case in Section 
\ref{sec:CPC} and the $CP$ violating one in Section \ref{sec:CPV}, which 
includes general considerations, the application to the specific case we 
are interested in, and leptogenesis. Finally, Section \ref{sec:concl} 
summarizes and concludes this work.

\section{\label{sec:frame}Framework}
The light neutrino Majorana mass matrix $m_\nu$ is given by 
\be \label{eq:mnu}
m_\nu = U^\ast \, \, m_\nu^{\rm diag} \, U^\dagger~.
\ee
Here $m_\nu^{\rm diag}$ is a diagonal matrix containing the 
light neutrino mass eigenstates $m_i$. 
Mixing is described by $U$, the unitary 
Pontecorvo--Maki--Nagakawa--Sakata \cite{PMNS} lepton 
mixing matrix, which can be parametrized as 
\bea \label{eq:Upara}
U = \left( \bad 
c_{12} c_{13} & s_{12} c_{13} & s_{13} e^{-i \delta} 
\\[0.2cm] 
-s_{12} c_{23} - c_{12} s_{23} s_{13} e^{i \delta} 
& c_{12} c_{23} - s_{12} s_{23} s_{13} e^{i \delta} 
& s_{23} c_{13} \\[0.2cm] 
s_{12} s_{23} - c_{12} c_{23} s_{13} e^{i \delta} 
& 
- c_{12} s_{23} - s_{12} c_{23} s_{13} e^{i \delta} 
& c_{23} c_{13}\\ 
               \ea   \right) 
 {\rm diag}(1, e^{i \alpha}, e^{i (\beta + \delta)}) 
\, , 
\eea
where $c_{ij} = \cos\theta_{ij}$, $s_{ij} = \sin\theta_{ij}$ and 
three physical phases $\alpha, \beta$ and $\delta$ are present. 
We shall work 
throughout this note in a basis in which the charged 
lepton mass matrix $m_\ell$ is real and diagonal.\\

\noindent The ranges of values of
the three neutrino 
mixing angles, which are allowed 
at 1(3)$\sigma$ by the current 
solar and atmospheric neutrino data and by the data
from the reactor antineutrino experiments
CHOOZ and KamLAND, read \cite{sandhya,bari}:
\be 
\label{eq:data}
\ba
0.35~(0.27) \leq \tan^2 \theta_{\rm sol} \equiv 
\tan^2 \theta_{12} \leq 0.52~(0.72)~,\\[0.3cm]
|U_{e3}|^2 = \sin^2\theta_{13} <  0.029~(0.074)~, \\[0.2cm]
\sin^2 2 \theta_{\rm atm} \equiv \sin^2 2 \theta_{23} \geq 0.95~(0.85)~.
\ea
\ee
Zero $U_{e3}$ and maximal atmospheric neutrino mixing are still 
allowed by the data, whereas maximal solar neutrino mixing is ruled out by 
more than 5$\sigma$ \cite{SNO2,sandhya}. 

\noindent For bimaximal neutrino mixing \cite{bimax} one has 
$\theta_{12} = \theta_{23} = \pi/4$ and $\theta_{13} = 0$. 
The corresponding neutrino mass matrix in case of $CP$ conservation 
then reads:   
\bea \label{eq:mnubimax}
m_\nu^{\rm bimax} = \left( \bad 
\D A & B & - B \\[0.2cm] 
\D \cdot & \D D + \frac{A}{2} &  \D D - \frac{A}{2} \\[0.3cm]
\D \cdot & \cdot & \D D + \frac{A}{2} 
\ea   \right)~, 
\eea
where 
\be \label{eq:ABD}
A = \frac{m_1^0 + m_2^0 }{2}~,~
B = \frac{m_2^0 - m_1^0}{2 \, \sqrt{2}}~
,~D = \frac{m_3^0 }{2}~. 
\ee
In case of $CP$ conservation, different relative signs of the mass states 
$m_i^0$ are possible.\\

\noindent 
On the theoretical side, the most general form  
\cite{seesaw,gora} for the light neutrino mass matrix is 
the type II see--saw formula 
\be \label{eq:typeII}
m_\nu = m_L - m_D^T \, M_R^{-1} \, m_D  \equiv m_\nu^{II} + m_\nu^I~,
\ee
where the second term $m_\nu^I$ 
is the conventional ordinary see--saw term. 
The first term is absent or suppressed 
in the major part of the theoretical works dedicated to neutrino physics. 
It can arise, for instance, in a large class of $SO(10)$ models in which the 
$B-L$ symmetry is broken by a {\bf 126} Higgs field \cite{so10typeII}. 
The relevant Lagrangian, which gives rise to the mass matrix 
(\ref{eq:typeII}) is 
\be \label{eq:L}
{\cal L} = \frac{1}{2}~(\nu_L^T)~C^{-1}~m_L~\nu_L 
+ \frac{1}{2}~(N_R^T)~C^{-1}~M_R~N_R  + 
\overline{N_R}~m_D~\ell_L + \overline{\ell_L}~m_\ell~\ell_R 
+ h.c.,
\ee
where $N_R$ ($\nu_L$) are the right--handed (left--handed) 
Majorana neutrinos and $\ell_{L,R}$ the charged leptons.  
In recent time, the situation in which the $m_L$ term dominates $m_\nu$ 
has gathered some attention because in certain $SO(10)$ models 
atmospheric neutrino mixing is a natural outcome of this dominance 
\cite{so10typeII}. 

\noindent 
In usual left--right symmetric theories 
$m_L$ and $M_R$ are proportional to  
the vacuum expectation values (vevs) of the 
electrically neutral components of scalar Higgs triplets, i.e., 
$m_L = f_L~v_L$ and $v_R = f_R~v_R$, where $v_{L,R}$ denotes the vevs and 
$f_{L,R}$ are symmetric $3 \times 3$ matrices. 
By acquiring the vev $v_R$, breaking of 
$SU(2)_L \times SU(2)_R \times U(1)_{B - L}$ 
to $SU(2)_L \times U(1)_Y$ is achieved \cite{gora}. 
The left--right symmetry 
demands the presence of both $m_L$ and $M_R$, and in addition 
it holds $f_R = f_L $. Whether $m_\nu^I$ or $m_\nu^{II}$ dominates 
$m_\nu$ depends on the details of the model.
 
\noindent Actually, in the class of models we refer to, 
there is a Higgs bi--doublet which gives rise to two 
Dirac mass terms. We shall assume for simplicity that only one of 
them plays a role for the lepton masses and later on for leptogenesis. 
This will be possible if, e.g., the coupling of one of the 
two $SU(2)_L$ doublets in the bi--doublet is suppressed by a 
very small vev.
 
\noindent Since left--right symmetry implies $f_R = f_L \equiv f$,  
the following relations hold: 
\be
\label{eq:prop}
m_L = \frac{v_L}{v_R} \, M_R~\mbox{ with } v_R \, v_L = \gamma \, v^2~, 
\ee
where $v$ is the SM Higgs vev and $\gamma$ a model dependent 
factor of order one. Thus, 
the neutrino mass matrix can be written as 
\be \label{eq:mnuII}
m_\nu = v_L \left( f - \frac{1}{\gamma \, v^2}~m_D^T~f^{-1}~m_D \right)~.
\ee 
The relation Eq.\ (\ref{eq:prop}) ensures that 
the type II see--saw mechanism \cite{gora} works, which sets 
the scale of neutrino masses by the vev $v_L$, 
where $v_L \propto 1/v_R$ with $v_R \gg m_D \gg v_L$. 
See \cite{ma} for a recent review and further references. 
Knowing the eigenvalues $m_i^0$ of $m_L$ will give the heavy Majorana 
neutrino masses $M_{i}$ through the relation 
\be \label{eq:masses}
M_{i} = \frac{v_R}{v_L} \, m_i^0 \simeq 3 \cdot 10^{10} \, 
\frac{1}{\gamma} \left(\frac{v_R}{10^{15} \, \rm GeV} \right)^2 \, 
\left(\frac{m_i^0}{10^{-3} \, \rm eV} \right) \, {\rm GeV}~.
\ee
In particular, the spectrum of the light Majorana masses in $m_L$ is 
identical to the spectrum of the heavy ones in $M_R$. The absence of 
$m_\nu^I$ would mean that the light neutrinos $\nu_i$, whose oscillations are 
currently measured, would have masses $m_i^0$. The mass 
spectrum of those neutrinos would then fix the mass spectrum of 
the heavy ones. See, e.g., \cite{WRlep,sche} for such a situation. In the 
remainder of the paper we shall discuss a small contribution of 
$m_\nu^I$ to $m_\nu$ and thus the spectrum of the light (oscillating) 
neutrinos will be slightly different from the one of the heavy Majorana 
neutrinos.\\  

\noindent Now we shall assume that 
the triplet term $m_L$ in Eq.\ (\ref{eq:typeII}) corresponds to 
$m_\nu^{\rm bimax}$. In order to obtain the physical light neutrinos 
one has to calculate the inverse matrix of $M_R$; 
using Eq.\ (\ref{eq:prop}) 
and $m_L = m_\nu^{\rm bimax}$ 
one finds: 
\bea 
M_R^{-1} = \frac{\D v_L}{\D v_R}~m_L^{-1} = 
\frac{\D v_L}{\D v_R}~
\left( \bad 
\tilde{A} & \tilde{B} & -\tilde{B}  \\[0.3cm]
\cdot & \tilde{D} + \frac{\D \tilde{A}}{\D 2} 
& \tilde{D} - \frac{\D \tilde{A}}{\D 2}\\[0.3cm] 
\cdot & \cdot & \tilde{D} + \frac{\D \tilde{A}}{\D 2} \\[0.3cm] 
\ea \right)~,
\eea 
where 
\be
\tilde{A} = \frac{A}{A^2 - 2 B^2}~,~ 
\tilde{B} = \frac{B}{2 B^2 - A^2}~,~
\tilde{D} = \frac{1}{4 D}~. 
\ee
Thus, $M_R^{-1}$ has a similar structure as $m_L$ or $m_\nu^{\rm bimax}$ 
in Eq.\ (\ref{eq:mnubimax}). 
We shall assume in the following a normal hierarchical mass spectrum, i.e., 
$(m_3^0)^2 \gg (m_{1,2}^0)^2$. This has the advantage that 
radiative corrections have only weak effects \cite{radcor}.

\noindent Now suppose that in the type II see--saw formula the term 
$m_L$ dominates. Furthermore, $m_D$ shall be hierarchical, i.e., 
\be
m_D \simeq {\rm diag}(0,0,m)~,
\ee
which is a very natural assumption, since $m_D$ is expected to be 
connected to the known fermion masses, which all display a very 
hierarchical mass spectrum. 
Then, from Eq.\ (\ref{eq:typeII}) one finds that the effect of the 
conventional see--saw term is a small contribution to the 
33 entry of $m_L$ given by 
\be \label{eq:s} 
s \equiv 
(m_D^T \, M_R^{-1} \, m_D)_{33} \simeq v_L^2~\frac{m^2}{4 \gamma \, v^2} 
\left(\frac{1}{m_1^0} + \frac{1}{m_2^0} + 
\frac{2 }{m_3^0} \right)~.
\ee
It is the main observation of the present study   
that this term can 
generate the observed sizable deviation from maximal solar neutrino mixing 
while at the same time also pulling $\theta_{\rm atm}$ and $U_{e3}$ away 
from their extreme ``bimaximal values'' $\pi/4$ and zero, 
respectively\footnote{Recently, 
it has been found that by adding to a conventional 
type I see--saw term a triplet contribution which is proportional 
to the unit matrix, one can promote a hierarchical mass spectrum to 
a partially degenerate one \cite{antusch}. 
The approach presented 
here is different and focuses on the mixing angles.}. 
 
\noindent The possible importance of this term in the class of models under 
consideration has been noted for the first time in \cite{eap99}.  
Typically, in these $SO(10)$ inspired models, it holds that $m \simeq v$, 
i.e., the Dirac mass matrix is related to the up--quark mass matrix. 
For smaller Dirac masses, i.e., when $m_D$ is related to the 
down quarks or charged leptons, $m_L$ receives negligible corrections from 
the conventional see--saw term \cite{WRlep}. If however 
$m \simeq v$, then we can estimate this term as 
\be \label{eq:sapp}
s \simeq \frac{0.1}{4 \gamma} \left( \frac{v_L}{10^{-2}~ \rm eV} \right)^2 
\left( \frac{10^{-3}~ \rm eV}{m_1^0} \right) ~{\rm eV}~, 
\ee 
where again hierarchical $m_i^0$ were assumed. A similar formula will 
hold when an inverted hierarchy with a non--zero smallest mass $m_3^0$ is 
present. 
Also, many non--singular mass matrices $M_R$ with hierarchical 
Dirac mass matrix will have the 33 entry as the leading term and can be cast 
in the form (\ref{eq:sapp}). For definiteness and the sake of simplicity, 
we shall stick to our left--right symmetry induced 
relation $m_L \propto M_R$. 
Naturally, for our 
reference values $m_1^0 = 10^{-3}$ eV and $v_L = 10^{-2}$ eV,  
the order of $s$ can be --- without varying $\gamma$ within more than 
one order of magnitude --- 
given by the scale of 
neutrino masses $\sqrt{\dms}$ or $\sqrt{\dma}$. 
The same is true for the case when the first two terms in 
Eq.\ (\ref{eq:s}) cancel each other, which will be of interest later on. 
Then it holds $s \simeq 0.01/(2\gamma) \, (v_L/10^{-2} \, {\rm eV})^2 \, 
(10^{-2} \, {\rm eV}/m_3^0)$ eV.  

\noindent Realistic Dirac mass matrices contain of course more than 
one non--vanishing entry. It is useful to parametrize $m_D$ in terms of 
a small parameter $\epsilon_D$, e.g., 
\be \label{eq:md}
m_D \simeq m 
\left( 
\bad 
0 & a \, \epsilon_D^3 & 0 \\[0.3cm] 
b \, \epsilon_D^3 & \epsilon_D^2 & c \, \epsilon_D^2 \\[0.3cm] 
0 & d \, \epsilon_D^2 & 1 
\ea
\right)~,
\ee
where $a, b, c, d$ are of order one and we may take 
$\epsilon_D \simeq 0.07$ in order to reproduce a realistic up--quark 
mass ratio with this matrix. 
The conventional see--saw term implied by this Dirac mass matrix 
will have the following leading form:
\be \label{eq:mdapp}
m_D^T ~M_R^{-1}~m_D \sim s 
\left( 
\bad 
\epsilon_D^6 & \epsilon_D^5 & \epsilon_D^3 \\[0.2cm]
\cdot & \epsilon_D^4 & \epsilon_D^2 \\[0.2cm]
\cdot & \cdot & 1 
\ea
\right)~,
\ee
which has only little effect on the results to be obtained for 
$m_D \simeq {\rm diag}(0,0,m)$.

\section{\label{sec:CPC}$CP$ conservation}

\subsection{General case}
We can easily 
diagonalize the mass matrix (\ref{eq:mnubimax}) when the term $s$ 
from Eq.\ (\ref{eq:s}) is subtracted from its 33 entry 
and the magnitude of $s$ is of order $D$ 
or smaller. One finds for the masses
\be \label{eq:CPCmass}
m_3 \simeq 2 D - \frac{s}{2}~,~m_{1,2} \simeq A - \frac{s}{4} \mp 
\sqrt{2 \, B^2 + \frac{s^2}{16}}~,
\ee
which for $s \rightarrow 0$ reproduces $m_i = m_i^0$. 
Using Eq.\ (\ref{eq:masses}), one could calculate the heavy Majorana 
neutrino masses. 
The atmospheric and solar $\Delta m^2$ read 
\bea \label{eq:CPCdm2}
\dms = m_2^2 - m_1^2 \simeq \sqrt{2B^2 + \frac{\D s^2}{\D 16}}~(4A - s)~, 
\\[0.2cm]
\dma = m_3^2 - m_2^2 \simeq 4D^2 - (A^2 + 2B^2) - 
s~\left(2 D - A - \frac{\D s}{\D 8} \right) - 
\frac{1}{2}~\dms~.
\eea
The mixing angles are given by 
\bea \label{eq:CPCmix}
\tan 2 \theta_{23} \simeq \frac{\D 2D - A}{\D s}~,~
\tan 2 \theta_{12} \simeq 4\sqrt{2} \, \frac{\D B}{\D s}~
\left(1 + \frac{\D s}{\D 2D - A} \right),~\\[0.2cm]
\tan 2 \theta_{13} \simeq \sqrt{2} \, \frac{\D B}{\D s}~
\frac{\D s}{\D 2D - A}\frac{\D s}{\D 2D - A - s/2}~.
\eea  
Bimaximal mixing is obtained --- as it should be --- for $s = 0$. 
Non--maximal solar neutrino mixing implies automatically non--maximal 
atmospheric neutrino mixing and non--zero $U_{e3}$.  
From the expression for $\theta_{12}$ and assuming hierarchical 
$m_{i}^0$ one obtains that $|s| \sim |m_2^0| \sim \sqrt{\dms}$ in order 
to reproduce the observations. 
From 
the expression for $\theta_{23}$ and assuming again hierarchical 
$m_{i}^0$ it follows that 
$m_3^0 \simeq \sqrt{\dma} \gs 2.4 \, (3, 4.4) \, s$ for 
$\sin^2 2 \theta_{23} > 0.85 \,(0.9, 0.95)$. 
With these values \dma{} is basically not affected by the conventional  
see--saw term. Also it holds that $m_3 \simeq m_3^0$ and hence 
the mass of the heaviest Majorana neutrino is proportional 
to the mass of the heaviest of the light physical neutrinos: 
$M_3 \simeq v_R/v_L \, m_3 \simeq  v_R/v_L \, \sqrt{\dma}$. 
The lightest heavy Majorana masses $M_{1,2}$ show however sizable 
deviations from $v_R/v_L \, m_{1,2}$. 

\noindent 
A useful parameter to describe the deviations from bimaximal mixing can be 
defined via \cite{WR} $U_{e2} = \sqrt{1/2}~(1 - \lambda)$, where 
$\lambda \simeq 0.22$ for typical best--fit points. In our framework one 
finds from the above expression for $\tan 2 \theta_{12}$ 
that to leading order $\lambda \simeq \sqrt{2}/16~s/B$. The other 
two deviations from bimaximal mixing can be described via 
Eq.\ (\ref{eq:devs}); the appropriate powers of $\lambda$ are 
$U_{e3} \simeq A_\nu~\lambda^4$ and 
$|U_{\mu_3}| \simeq \sqrt{1/2}~(1 - B_\nu~\lambda^2)$, where $A_\nu$ and 
$B_\nu$ are of order one and are functions of $A, B$ and $D$. 
Note that $\lambda \propto s$, quantifying again that it is 
the small contribution $s$ of the conventional see--saw term that is 
responsible for the deviation from bimaximal mixing. \\ 

\noindent Eq.\ (\ref{eq:CPCmix}) and the 
fact that $2D \gg A, s/2$ can be used to obtain an interesting 
correlation of neutrino mixing observables: 
\be \label{eq:CPCsum}
|U_{e3}| \simeq \frac{1}{4} \, 
\frac{\tan \theta_{12}}{1 - \tan^2 \theta_{12}} \, 
\frac{1 - \sin^2 2\theta_{23}}{\sin^2 2\theta_{23}} 
~.
\ee
The larger the deviation from maximal atmospheric neutrino mixing 
and the larger $\tan^2 \theta_{12}$, the more sizable becomes 
$U_{e3}$.

\noindent Note that, though the mixing angles and the 
mass states are altered, 
the $ee$ entry of $m_\nu$ (the so-called effective Majorana mass),  
which is in principle 
measurable in neutrinoless double beta--decay experiments \cite{0vbbrev}, 
is not changed by the procedure. We can express, however, the effective mass 
$\meff \equiv A$ through $s$ and the neutrino mixing parameters, e.g., 
\be \label{eq:meffNH}
\meff \simeq   \sqrt{\dma} - 
s \frac{\sin 2 \theta_{23}}{\sqrt{1 - \sin^2 2 \theta_{23}}}~.
\ee
Both terms can be sizable, but are subtracted so that small 
$\meff \sim m_2^0 \sim \sqrt{\dms}$ is obtained.

\begin{figure}
\begin{center}
\epsfig{file=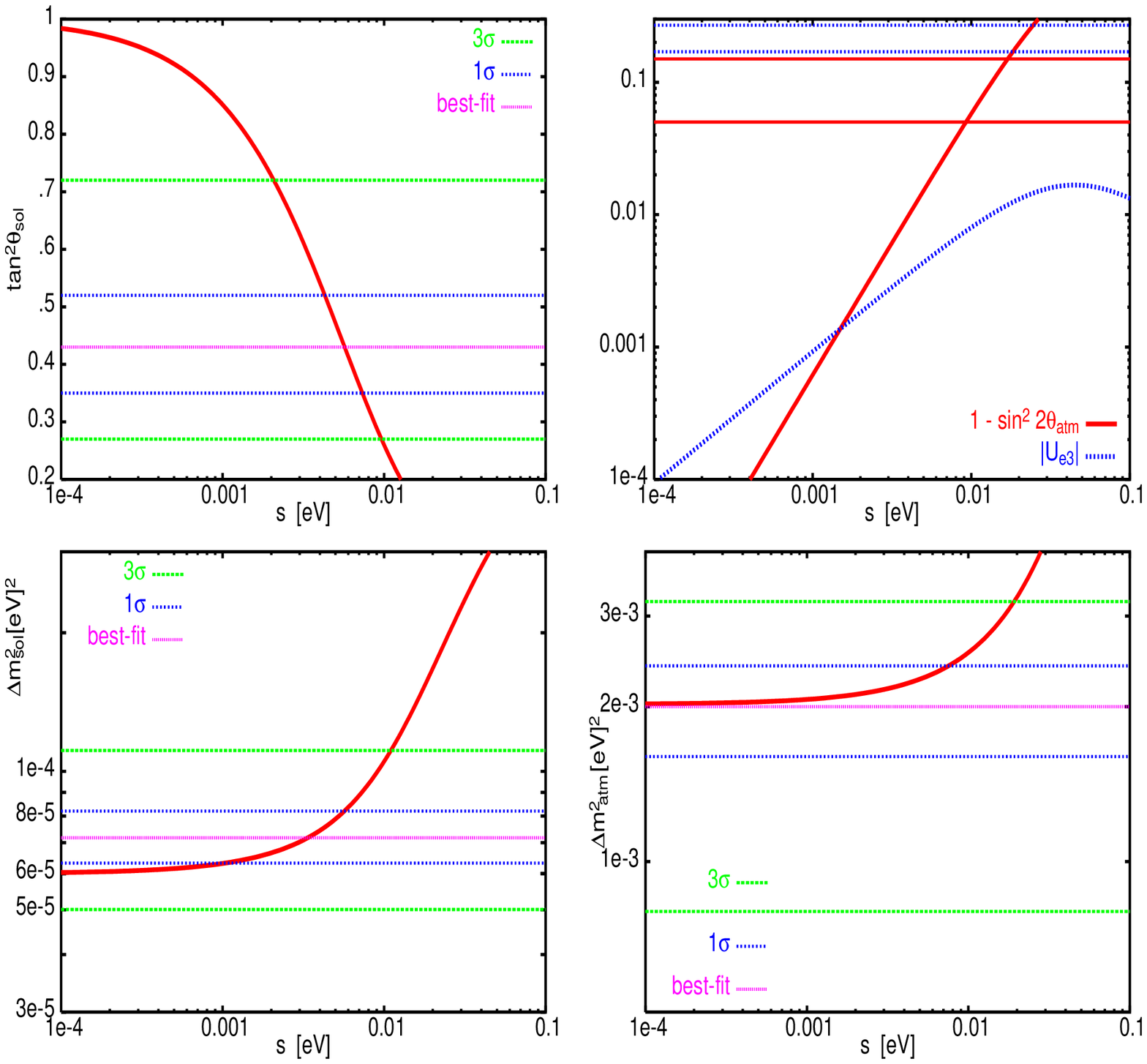,width=18cm,height=25cm}
\vspace{-5cm}
\caption{\label{fig:CPC}Neutrino mixing parameters as a function of $s$ in 
case of $CP$ conservation.} 
\end{center}
\end{figure}

\noindent In Fig.\ \ref{fig:CPC} we show 
the neutrino mixing parameters 
as a function of $s$ obtained with the Dirac mass matrix $m_D$ 
from Eq.\ (\ref{eq:md}), where for simplicity we set the parameters 
$a = b = c = d = 1$. For the eigenvalues of $m_L$ we choose 
$m_3^0 = 0.045$ eV, $m_2^0 = 0.008$ eV and $m_1^0 = 0.002$ eV. 
Indicated are the best--fit points as well as the 
1 and 3$\sigma$ ranges of the oscillation parameters\footnote{For the 
atmospheric neutrino parameter a novel (unpublished) analysis of 
the SuperKamiokande collaboration yields a best--fit value of 
$\dma = 2.0 \cdot 10^{-3}$ eV$^2$ \cite{SKaachen}. 
We take as 1 (3)$\sigma$ errors 
0.4 $(1.2) \cdot 10^{-3}$ eV$^2$, which are the errors obtained in 
an earlier analysis \cite{bari}.}. 
With Eqs.\ (\ref{eq:CPCdm2}) and (\ref{eq:CPCmix}) one finds that for, e.g., 
$s = 0.005$ eV the results are $\dma \simeq 2.2 \cdot 10^{-3}$ eV$^2$, 
$\dms \simeq 8.1 \cdot 10^{-5}$ eV$^2$, $\tan^2 \theta_{12} \simeq 0.48$, 
$1 - \sin^2 2 \theta_{23} \simeq 0.015$ and $|U_{e3}| \simeq 0.0044$. 
Good agreement between these numbers and the plots is found. 
From the figure it is seen that values of $s$ of a few times $10^{-3}$ 
eV reproduce the observed deviation from maximal solar neutrino mixing, 
while predicting small 
$U_{e3}$ of few times $10^{-3}$ and $1 - \sin^2 2 \theta_{23}$ around 
few times $10^{-2}$.  
For values of $s \gs 0.01$ eV the parameters except for $|U_{e3}|$ 
leave their experimentally allowed ranges. 

\noindent Unfortunately, the implied value of $|U_{e3}|$ is too small to be 
measured in the next future. Values of this parameter below 
0.01 are probably only accessible by a neutrino factory \cite{nufac}. 
The implied values of $1 - \sin^2 2 \theta_{23} \sim 0.01$, 
however, could be testable by next generation long--baseline experiments 
such as JHF--SK \cite{JHF} or NuMI off--axis \cite{off}, 
all of which claim a sensitivity of 
$\sigma(\sin^2 2 \theta_{23}) \simeq 0.01$.

\subsection{Generating \dms}
\noindent It is obvious from Eqs.\ (\ref{eq:mnubimax}) and (\ref{eq:ABD})  
that for $A = 0$ one would start with vanishing\footnote{This is 
in principle also possible when we set $B = 0$. However, as seen from 
Eq.\ (\ref{eq:mnubimax}), a neutrino mass matrix with zeros in the 
$e\mu$ and $e\tau$ entry would result, which is known not to 
reproduce neutrino data \cite{FGM}. From Eq.\ (\ref{eq:CPCmix}) one 
finds in this particular example that solar neutrino mixing would vanish.} 
$\dms$. 
Then, after adding the conventional see--saw term $s$, 
the induced mass squared difference reads 
(see Eq.\ (\ref{eq:CPCdm2})):  
\be \label{eq:dmsA0}
\dms \simeq s~\sqrt{2 B^2 + \frac{s^2}{16}} \simeq  
\frac{s^2}{4} \frac{1 + \tan^2 \theta_{12}}{1 - \tan^2 \theta_{12}}~,
\ee
i.e., the small conventional see--saw contribution can not only 
describe the deviation from maximal solar neutrino mixing but also 
induce non--zero $\dms \ll \dma$. 
Inserting the best--fit points 
of $\dms = 7.17 \cdot 10^{-5}$ eV$^2$ and 
$\tan^2 \theta_{12} = 0.43$ \cite{sandhya} in the last equation yields 
that $s \simeq 0.01$ eV, i.e., a value a bit larger 
than in the $A \neq 0$ case. Consequently, 
both $U_{e3}$ and $1 - \sin^2 2 \theta_{23}$ 
will also be somewhat larger. For instance, from 
Eq.\ (\ref{eq:meffNH}) one can get 
\be 
\sqrt{\dma} \simeq 
s~\frac{\sin 2 \theta_{23}}{\sqrt{1 - \sin^2 2 \theta_{23}}}~, 
\ee
which for $\dma = 2 \cdot 10^{-3}$ eV$^2$ and $s \simeq 0.01$ eV 
yields $1 - \sin^2 2 \theta_{23} \simeq 0.05$. Another way to 
distinguish the cases $A = 0$ and $A \neq 0$ would be to note that 
$A = \meff$ and then to prove that $\meff = 0$. This, however,  
will in practice not be possible. 

\noindent It is now for $A = 0$ also possible to give a concise formula for 
the phenomenologically interesting ratio of the solar and atmospheric 
mass squared ratios. Using Eq.\ (\ref{eq:CPCdm2}) one finds 
\be
\frac{\dms}{\dma} \simeq \frac{1}{4} \, 
\frac{1 + \tan^2 \theta_{12}}{1 - \tan^2 \theta_{12}} \, 
\frac{1 - \sin^2 2\theta_{23}}{\sin^2 2\theta_{23}}~.
\ee
The ratio of the mass squared differences is thus linked to a 
small deviation of $\theta_{23}$ from $\pi/4$. 

\noindent Note that from Eqs.\ (\ref{eq:CPCdm2}) and (\ref{eq:dmsA0}) 
it holds that $R \propto s^2$. Remembering from above that 
the parameter describing the deviations from bimaximal mixing is 
$\lambda \propto s$ we see that $R \propto \lambda^2$. Therefore, 
the framework described here gives an explanation for the fact that 
the ratio of \dms{} and \dma{} is numerically linked \cite{WR} to the observed 
deviation from maximal solar neutrino mixing.  

\begin{figure}
\begin{center}
\epsfig{file=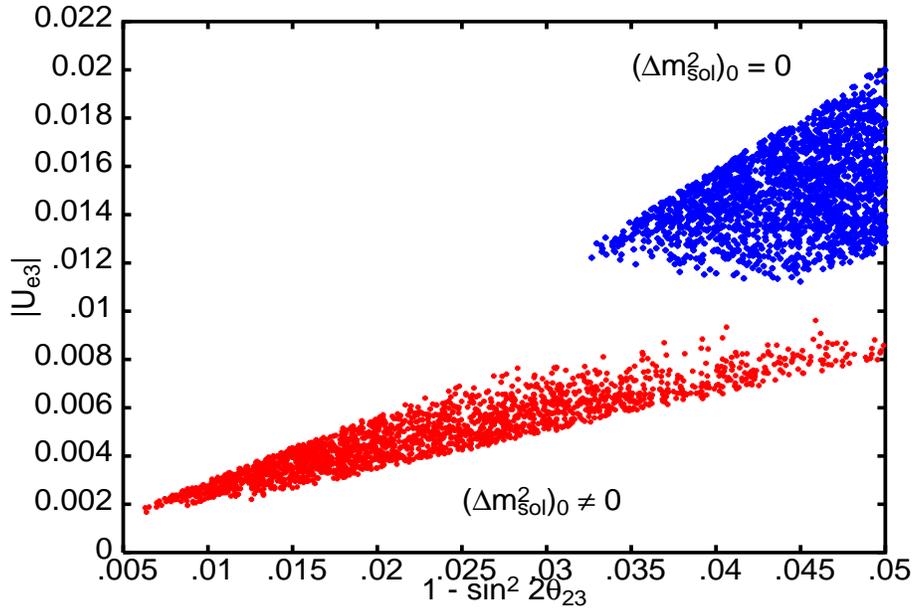,width=12cm,height=8cm}
\caption{\label{fig:CPCNHsc}Scatter plot of the observables $|U_{e3}|$ and 
$1 - \sin^2 2 \theta_{23}$ obtained for the cases $A=0$ 
and $A \neq 0$ in case of $CP$ conservation.} 
\end{center}
\end{figure}

\noindent In Fig.\ \ref{fig:CPCNHsc} we 
show a scatter plot of the observables $|U_{e3}|$ and 
$1 - \sin^2 2 \theta_{23}$ obtained for the cases $A=0$ 
and $A \neq 0$. To produce the plots, $m_i^0$ was varied according to a 
hierarchical spectrum and $s$ was required to be smaller than 
$m_3^0$. 
The oscillation parameters were required to lie inside 
their 1$\sigma$ ranges. Both parameters are seen to prefer larger 
values when \dms{} is induced by the type I see--saw term. 
In this particular example and if 
$A \neq 0$ ($A = 0$) lower limits on $U_{e3}$ 
of 0.002 (0.011) can be set. For $1 - \sin^2 2 \theta_{23}$ the lower 
limits is 0.006 (0.033). 
Though being slightly larger, the indicated values of 
$|U_{e3}| \gs 0.01$ mean that they are still 
too small for next generation experiments but testable by the 
JHF--HK setup \cite{JHF}.

\section{\label{sec:CPV}$CP$ violation}

\subsection{\label{sec:CPVgen}General considerations}
In type II see--saw models the number of independent phases 
is obviously larger than in type I. It has been shown \cite{jose1,phaII} that 
the Lagrangian (\ref{eq:L}) contains (for 3 left-- and 3 right--handed 
neutrinos) 12 independent phases\footnote{Suppose both $m_L$ and $M_R$ are 
real and diagonal. Then any $CP$ violation will stem from 
$m_D$ and $m_\ell m_\ell^\dagger$, which possess in total 
12 phases \cite{phaII}.}. Let us write the relevant matrices $f$ and 
$m_D$ in the following way:
\be \label{eq:fCPV}
f = U_f^\ast~f^{\rm diag}~U_f^\dagger~\mbox{ and }~
m_D = U_1^\dagger~m_D^{\rm diag}~U_2~.
\ee
The eigenvalues of $m_L$ are given by $m_i^0 = v_L ~f^{\rm diag}_i$.  
Unitary matrices such as $U_f$ can always be written as \cite{PPR} 
\be \label{eq:PPR}
U_f = e^{i \phi}~P_f~\tilde{U}_f~Q_f~,
\ee
where $P_f={\rm diag}(1,e^{i \alpha},e^{i \beta})$ and 
$Q_f={\rm diag}(1,e^{i \rho},e^{i \sigma})$ are diagonal matrices 
containing 2 phases each and $\tilde{U}_f$ is a unitary matrix 
parametrized in analogy to the CKM matrix, i.e., it is defined by 
3 angles and 1 phase. Analogous definitions hold for $U_1$ and $U_2$. 
Using Eq.\ (\ref{eq:mnuII}) one finds after 
some simple steps: 
\begin{small}
\bea  \label{eq:mnugen}
m_\nu = v_L~\left( e^{-2i \phi} P_f^\dagger~\tilde{U}_f^\ast~Q_f^\dagger~
f^{\rm diag}~Q_f^\dagger~\tilde{U}_f^\dagger~P_f^\dagger -
\right. \\[0.2cm]
\left.  \frac{\D e^{i(2\phi_2 - 2\phi_1 + \phi)}}{\D \gamma \, v^2}~
Q_2~\tilde{U}_2^T~\tilde{P}_2~m_D^{\rm diag}~\tilde{U}_1^\ast~Q_1^\dagger
~P_f~\tilde{U}_f~Q_f^{2}~(f^{\rm diag})^{-1}~
\tilde{U}_f^T~P_f~Q_1^\dagger~\tilde{U}_1^\dagger~
m_D^{\rm diag}~\tilde{P}_2~\tilde{U}_2~Q_2 
\right),
\eea 
\end{small}
where 
$\tilde{P}_2 = P_1^\dagger~P_2$ was defined. 
In the neutrino mass term $\nu^T \, m_\nu \, \nu$ one can 
choose now new fields 
$\nu \rightarrow \nu^\prime \equiv e^{-i \phi}~P_f^\dagger \, \nu$, 
and make the same transformation for the charged leptons. 
This will alter the above equation to 
\begin{small}
\bea  \label{eq:mnugen1}
m_\nu = v_L~\left( \tilde{U}_f^\ast~Q_f^\dagger~
f^{\rm diag}~Q_f^\dagger~\tilde{U}_f^\dagger -
\right. \\[0.2cm]
\left.  \frac{\D e^{2i(\phi_2 - \phi_1 + \phi)}}{\D \gamma \, v^2}~
P_f~Q_2~\tilde{U}_2^T~\tilde{P}_2~m_D^{\rm diag}~\tilde{U}_1^\ast~Q_1^\dagger
~P_f~\tilde{U}_f~Q_f^{2}~(f^{\rm diag})^{-1}~
\tilde{U}_f^T~P_f~Q_1^\dagger~\tilde{U}_1^\dagger~
m_D^{\rm diag}~\tilde{P}_2~\tilde{U}_2~Q_2~P_f
\right)
\eea 
\end{small}
Thus, at this point, 
$m_\nu^{II}$ contributes with 3 phases (2 in $Q_f$ and 
1 in $\tilde{U}_f$) to $m_\nu$ 
and the conventional type I term with an additional 
9 (1 common, 2 in $P_f~Q_2$, 1 each in $\tilde{U}_{1,2}$, 2 each in $Q_1$ and 
$\tilde{P}_2$), corresponding to the 
mentioned result of 12 independent $CP$ restrictions. 
As two known limits, consider the cases when the second or the 
first term is absent. If the second term vanishes, 
we have nothing more to absorb and hence we 
end up with the well--known result of three physical phases for a symmetric 
$3 \times 3$ neutrino mass matrix.  
If only the conventional see--saw term is present and in addition 
the right--handed Majorana neutrino mass matrix is real and diagonal 
(as it is possible to go into this basis), 
we have $\tilde{U}_f = P_f = Q_f = \mathbbm{1}$. Then, from 
the second term in Eq.\ (\ref{eq:mnugen1}) the 2 phases in $Q_2$ and the 
common phase $ 2(\phi_2 - \phi_1 +  \phi)$ can be absorbed 
in the charged lepton fields and there are in total 6 phases, 
which will combine in a complicated manner to the three measurable ones. 
Six is the well--known number of independent $CP$ restrictions in general  
three neutrino type I see--saw models \cite{PPR,branco}.

\subsection{\label{sec:spec}Our special case}
Our requirement of $m_L$ 
being bimaximal will remove the phase from $\tilde{U}_f$ and the presence 
of 3 zeros (or very small entries) in $m_D$ will render 3 more phases 
unphysical. 
Let us define  
\be \label{eq:mdCPV}
m_D \simeq m 
\left( 
\bad 
0 & a \, e^{i \varphi_1} \, \epsilon_D^3 & 0 \\[0.3cm] 
b \, e^{i \varphi_2}\, \epsilon_D^3 & e^{i \varphi_3} \, \epsilon_D^2 
& c \, e^{i \varphi_4 } \, \epsilon_D^2 \\[0.3cm] 
0 & d \, e^{i \varphi_5} \, \epsilon_D^2 & e^{i \varphi_6} 
\ea
\right)~.
\ee
Then, the 33 entry of the conventional see--saw term reads 
\bea \label{eq:sCPV}
s = (m_D^T \, M_R^{-1} \, m_D)_{33}
\simeq v_L^2~\frac{\D m^2}{\D 4 \, \gamma \, v^2} \, 
e^{2i(\beta + \phi + \varphi_6)} 
\left( 
\frac{\D 1}{\D m_1^0} 
+ \frac{\D e^{2i\rho}}{\D m_2^0} + \frac{\D 2 \, e^{2i\sigma}}{\D m_3^0}
\right) \\[0.2cm]
\simeq 
v_L^2~\frac{\D m^2}{\D 4 \, \gamma \, v^2} 
\frac{\D e^{2i(\beta + \phi + \varphi_6)}}{\D m_1^0}
~. 
\eea
All other entries of $m_D^T \, M_R^{-1} \, m_D$ are suppressed by terms 
of at least order $\epsilon_D^2$. 
With the above $s$ the neutrino mass matrix is 
\be
m_\nu \simeq v_L 
e^{-2i\phi} ~P_f^\dagger~\tilde{U}_f^\ast~Q_f^\dagger~f^{\rm diag}~
Q_f^\dagger~\tilde{U}_f^\dagger~P_f^\dagger 
- |s| \, e^{2i(\beta + \phi + \varphi_6)}~{\rm diag}(0,0,1)
~.
\ee 
In the neutrino mass term $\nu^T \, m_\nu \, \nu$ one can  
choose now as before new fields according to 
$\nu \rightarrow \nu^\prime \equiv e^{-i \phi}~P_f^\dagger \, \nu$, 
and make the same transformation for the charged leptons. 
This will alter the above equation to 
\be \label{eq:CPVfin}
m_\nu \simeq v_L \,  
\tilde{U}_f^\ast~Q_f^\dagger~f^{\rm diag}~
Q_f^\dagger~\tilde{U}_f^\dagger 
- |s| \, 
e^{i \delta_I}
~{\rm diag}(0,0,1)
~,
\ee
where we have defined 
\be \label{eq:deltaI}
\delta_I \equiv 2(\varphi_6 + 2(\beta + \phi)) ~.
\ee
Recall that due to its bimaximality $\tilde{U}_f$ is real.   
The difference between the $CP$ violating and 
conserving cases is the presence of two ``Majorana--like'' phases 
$\rho, \sigma$ in $Q_f$ for the eigenvalues of $m_L$ and the 
subtraction of a small term $s$, which in 
general is now complex. The phase of $s$ is a combination of one phase in 
$m_D$ and two in $m_L$. Note that in general the parameter $\gamma$ 
appearing in $s$ could also be complex and therefore could 
also contribute to the relative phase between 
the two terms in Eq.\ (\ref{eq:CPVfin}).

\noindent It is interesting to consider $CP$ violating observables in the 
lepton sector. The rephasing invariant 
$J_{CP}$ \cite{JCP}, which governs the magnitude of 
$CP$ violating effects in neutrino oscillations \cite{SPKR},  
can be written in terms of the neutrino mass matrix as \cite{branco2} 
\be
J_{CP} = -\frac{{\rm Im} (h_{12} \, h_{23} \, h_{31} )}
{\Delta m^2_{21} \, \Delta m^2_{31} \, \Delta m^2_{32}}
~,  \mbox{ where } 
h = m_\nu m_\nu^\dagger ~. 
\ee
In case of $m_2^0 \neq m_1^0$, i.e., \dms{} not generated by the 
conventional see--saw term, 
one finds from Eq.\ (\ref{eq:CPVfin}) that for our chosen set of 
parameters the leading term is 
\be \label{eq:JCP1}
J_{CP} \simeq \frac{(m_2^0)^2 \, (m_3^0)^2 \, s}{16\,  (\dma)^2 \, \dms} 
\left( 
m_1^0 \, \sin(\delta_I + 2 \rho) - m_2^0 \, \sin \delta_I 
\right)
\simeq \frac{-(m_2^0)^3 \, (m_3^0)^2 \, s}{16\,  (\dma)^2 \, \dms} 
\sin \delta_I ~,
\ee
which, as it should, vanishes for $s = 0$ because this situation 
would correspond to exact bimaximal neutrino mixing. The order 
of magnitude is for $m_3^0 = 0.045$ eV, 
$m_2^0 = 0.008$ eV and $m_1^0 = 0.002$ eV and the best--fit 
values of the $\Delta m^2$ given by $|J_{CP}|^{\rm max} \simeq 10^{-3}$. 
Recall that in terms of neutrino mixing angles, 
\be
J_{CP} = \frac{1}{8} \sin 2 \theta_{12} \, \sin 2 \theta_{23} 
\, \sin 2 \theta_{13} \, \cos\theta_{13} \, \sin \delta 
\simeq \frac{1}{4}  \sin 2 \theta_{12} \, |U_{e3}| \, \sin \delta 
\sim \frac{|U_{e3}|}{4}\, \sin \delta ~.
\ee
We conclude that $|U_{e3}|$ is of order $10^{-3}\ldots 10^{-2}$. 
Note that the $ee$ element of the neutrino mass matrix, which is measurable  
in neutrinoless double beta decay, is given by 
\be \label{eq:meffCPV}
\meff = \frac{1}{2} \left( m_1^0 + e^{-2 i \rho} m_2^0 \right)~.
\ee
Therefore, in this particular case the $CP$ violation in neutrino 
oscillation as governed by 
$J_{CP}$, which depends only very weakly on $\rho$, is decoupled from 
the parameter which is responsible for cancellations 
in the effective mass governing neutrinoless double beta decay.\\[0.2cm]

\noindent Now let us consider the case when \dms{} is induced by the 
conventional see--saw term. From Eq.\ (\ref{eq:sCPV}) one finds that for 
$m_1^0 = m_2^0$ and $m_3^0 \gg m_2^0$:  
\be
s^\prime \simeq v_L^2~\frac{\D m^2}{\D 2 \, \gamma \, v^2} \, 
\frac{\cos \rho}{m_2^0} \, 
e^{i( \rho + 2(\beta + \phi + \varphi_6))} ~, 
\ee
i.e., a different phase than in Eq.\ (\ref{eq:deltaI}) is present. 
The leading term in $J_{CP}$ will for $m_3^0 \gg m_2^0$ be proportional 
to 
\be
J_{CP} \propto |s^\prime| \left( m_3^0 \, \cos (\delta_I + 2 \rho) - 
|s^\prime| \, \cos(\rho - 2 \sigma) \right) \, \sin \rho ~.
\ee
Since in this case (see Eq.\ (\ref{eq:meffCPV})) we have  
$\meff = m_2^0 \, \cos \rho$, we find a direct connection between 
the effective mass in neutrinoless double beta decay and $J_{CP}$. 

\noindent Regardless if $(\dms)^0 = 0$ or $(\dms)^0 \neq 0$, the entries 
$A = (m_1^0 + e^{-2i \rho} m_2^0)/2$ and 
$B = (m_1^0 - e^{-2i \rho} m_2^0)/\sqrt{8}$ should be of the same order 
to reproduce bi--large neutrino mixing. 
This implies that $\rho \sim \pi/4$, which is confirmed by 
a numerical analysis. 
Also for the $CP$ violating case under 
consideration, $1 - \sin^2 2\theta_{23}$ and $|U_{e3}|$  
will be significantly larger when $(\dms)^0 = 0$. 
Consequently, also $|J_{CP}|$ will be larger. 
Let us choose for a numerical analysis 
again the values $m_3^0 = 0.045$ eV, 
$m_2^0 = 0.008$ eV and $m_1^0 = 0.002$ eV or $m_2^0 = m_1^0 = 0.008$ eV 
when \dms{} is to be generated by $m_\nu^I$.   
One will observe that for $(\dms)^0 \neq 0$ ($(\dms)^0 = 0$) 
$|s| \ls 0.01$ eV ($|s^\prime| \gs 0.01$ eV) is the 
preferred value in order to reproduce the neutrino 
mixing data. Choosing for definiteness $|s| = 0.007$ eV 
and $|s^\prime|/\cos \rho = 0.015$ eV, respectively, we can analyze the 
requisite values of the other parameters. 

\begin{figure}
\begin{center}
\vspace{-3cm}
\epsfig{file=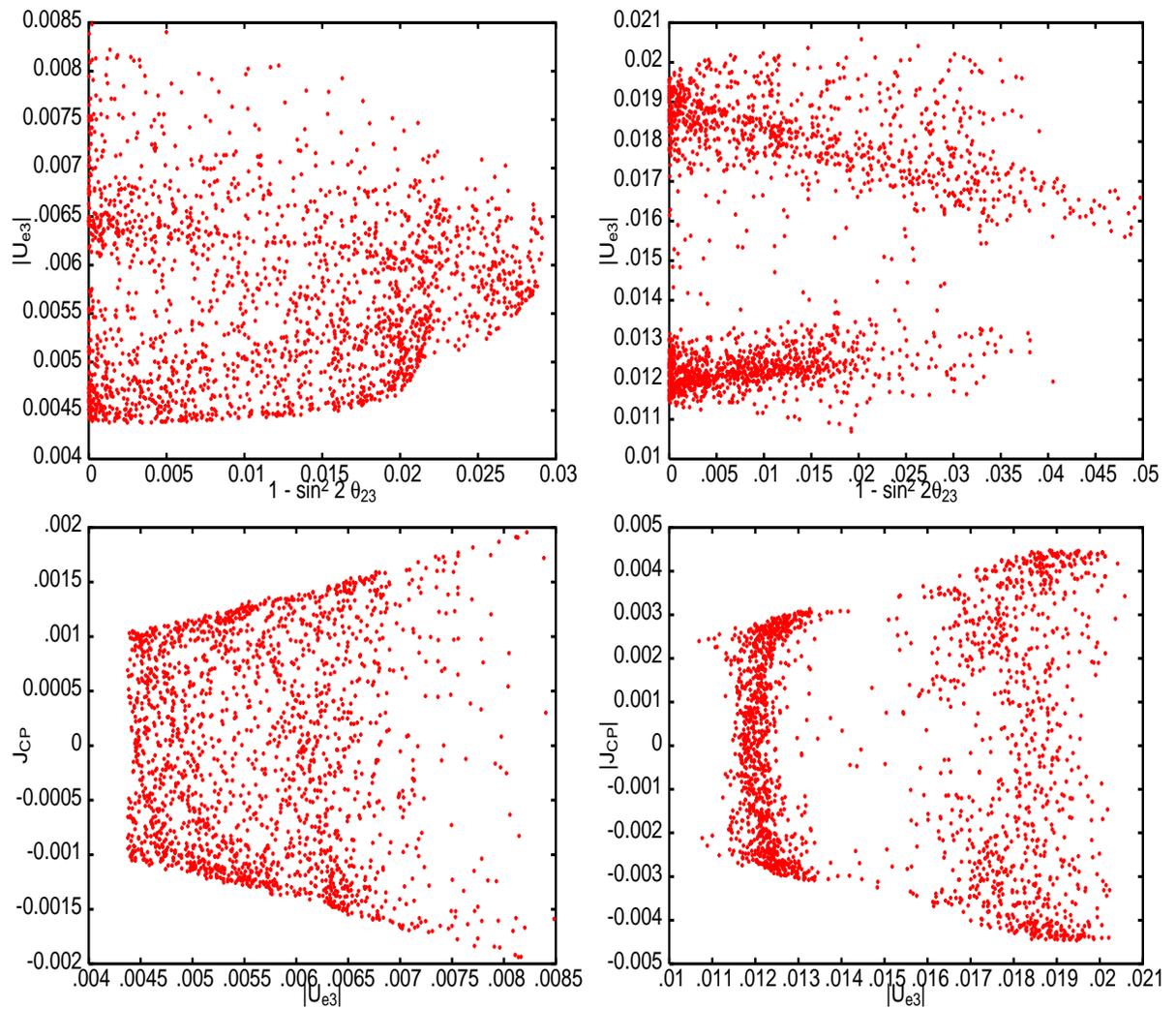,width=18cm,height=24cm}
\vspace{-1.5cm}
\caption{\label{fig:CPV}Scatter plots of  
neutrino mixing parameters $1 - \sin^2 2 \theta_{23}$ against $|U_{e3}|$ 
and $|U_{e3}|$ against $J_{CP}$ 
when $(\dms)^0 \neq 0$ 
(left column) and $(\dms)^0 = 0$ (right column).}
\end{center}
\end{figure}

\noindent In Fig.\ \ref{fig:CPV} 
we show the results for some important quantities, namely 
$1 - \sin^2 2 \theta_{23}$ against $|U_{e3}|$ and 
 $|U_{e3}|$ against $J_{CP}$. 
As mentioned before, also in the $CP$ conserving case the values of 
$1 - \sin^2 2 \theta_{23}$ and $|U_{e3}|$ can be significantly larger 
than $(\dms)^0 = 0$. Note, however, that now in case of $CP$ 
violation atmospheric neutrino mixing can be maximal. 
Non--zero $U_{e3}$ is however always guaranteed.

\subsection{\label{sec:lepto}Leptogenesis}
Leptogenesis \cite{lepto} in the framework of 
type II see--saw mechanism has so far not been discussed in as many 
details as the type I case (see, e.g., \cite{lepto03} and references 
therein).  
The presence of the Higgs triplet $\Delta_L$ implies the existence 
of novel decay processes capable of producing a decay asymmetry. 
In the usual type I see--saw approach the decay $N_i \ra \Phi \, \ell$, 
where $N_i$ is one of the heavy Majorana neutrinos, 
$\Phi$ the Higgs doublet and $\ell$ a lepton, 
receives 1--loop self--energy and vertex 
corrections, where for the latter a virtual heavy Majorana neutrino $N_j$ 
is exchanged. The decay asymmetry stemming from these two diagrams will 
be called $\varepsilon_1^N$. 
When a triplet is present, it also will be exchanged in the 
vertex correction to the decay $N_1 \ra \Phi \, \ell$ \cite{epsTalt,hamsen}, 
giving rise to a decay asymmetry  $\varepsilon_1^\Delta$. 
Furthermore, the decay $\Delta_L \ra  \ell \, \ell$ is possible, 
which will receive 1--loop vertex corrections via 
virtual Majorana neutrino exchange \cite{epsTalt,hamsen}. 
If the triplet mass $M_\Delta$ is much larger than the Majorana neutrino 
masses, the baryon asymmetry is produced via the decay of 
the Majorana neutrinos. Let us focus on this situation, since typically 
for the mass of the triplet $m_{\Delta_L} \sim v_R$ holds, which is 
larger than the mass of the lightest of the heavy Majorana neutrinos. 
The decay asymmetries for the heavy Majorana neutrino decay 
read 
\bea
\varepsilon_1^N = \frac{\D 1}{\D 8 \pi \, v^2} \, 
\frac{\D 1}{\D (\tilde{m}_D \tilde{m}_D^\dagger)_{11}} \, 
\sum\limits_{j = 2,3} \, 
{\rm Im} \left\{ (\tilde{m}_D \tilde{m}_D^\dagger)^2_{1j} \right\} \, 
f_N\left(\frac{\D M_j^2}{\D M_1^2} \right)~, \\[0.3cm]
\varepsilon_1^\Delta = \frac{\D -3 \, v_L}{\D 8 \pi \, v^2} \, 
\frac{\D 1}{\D (\tilde{m}_D \tilde{m}_D^\dagger)_{11}} \, 
\frac{\D M_\Delta^2}{\D M_1} \, 
{\rm Im} \left\{ (\tilde{m}_D f^\ast \, \tilde{m}_D^T)_{11} \right\} \, 
f_\Delta\left(\frac{\D M_\Delta^2}{\D M_1^2} \right)~,
\eea
where $\varepsilon_1^\Delta $ has been calculated recently 
\cite{hamsen,anki}.  
We wrote the expressions in terms of $\tilde{m}_D = U_f^\dagger \, m_D$, 
because we have to 
work in the basis in which the heavy Majorana neutrinos are diagonal. 
The functions $f_N$ and $f_\Delta$ are given by 
\bea
f_N(x) = \sqrt{x} \left( 1 - (1 + x) \log(1 + 1/x) 
+ \frac{\D 1}{\D 1-x} \right) \simeq \frac{\D -3}{\D 2 \sqrt{x}}~, \\[0.3cm]
f_\Delta (x) = 1 - x \log(1 + 1/x) \simeq \frac{\D 1}{\D 2 x}~,
\eea 
where the limits for $x \gg 1$ were given. Using these approximations, 
the asymmetries can be written as 
\bea 
\varepsilon_1^N = \frac{\D 3}{\D 16 \pi \, v^2} \, 
\frac{\D M_1}{\D (\tilde{m}_D \tilde{m}_D^\dagger)_{11}} \,  
{\rm Im} \left\{ (\tilde{m}_D (m_\nu^{I})^\ast \, \tilde{m}_D^T)_{11} \right\}
~, \\[0.3cm]
\varepsilon_1^\Delta = \frac{\D 3}{\D 16 \pi \, v^2} \, 
\frac{\D M_1}{\D (\tilde{m}_D \tilde{m}_D^\dagger)_{11}} \, 
{\rm Im} 
\left\{ 
(\tilde{m}_D (m_\nu^{II})^\ast \, \tilde{m}_D^T)_{11} 
\right\}~.
\eea
As first observed in \cite{hamsen}, if 
$m_\nu^I$ ($m_\nu^{II}$) dominates in $m_\nu$, one would expect 
$\varepsilon_1^N$ ($\varepsilon_1^\Delta$) to dominate the 
decay asymmetry. 
To check this assumption in our scenario, let us 
write  
$m_D \simeq m \, e^{i \varphi_6} \, {\rm diag}(0,0,1)$ and take 
$U_f$ and $f$ from Eq.\ (\ref{eq:fCPV}) and (\ref{eq:PPR}) in order 
to calculate $m_\nu^{II} = v_L \, U_f^\ast~f^{\rm diag}~U_f^\dagger$. 
For $m_\nu^{I}$ we have to calculate $M_R^{-1}$, which is given by 
$v_L/v_R \, m_L^{-1} = U_f~(v_R~f^{\rm diag})^{-1}~U_f^T$. 
The result for $m_3^0 \gg m_{1,2}^0$ and $m_2^0 \neq m_1^0$ is  
\bea  \label{eq:epss} 
\varepsilon_1^N \simeq \frac{\D -3 }{\D 64 \, \pi } 
\, \frac{\D m^2}{\D v^2} \, \frac{\D v_L}{\D v_R} \, \frac{\D M_1}{\D m_1^0}
\, \sin 4 \varphi_6 ~, 
\\[0.3cm]
\varepsilon_1^\Delta \simeq  
\frac{\D -3}{\D 32 \, \pi \, v^2} \, 
 M_1~m_3^0 ~\sin 2 (2\beta + 2 \phi - \varphi_6 + \sigma)
~,
\eea
We can clarify the situation significantly when we note that 
$v_L/v_R = v_L^2/(\gamma \, v^2)$ and by glancing at Eq.\ (\ref{eq:sCPV}), 
where $s$ is defined. Furthermore, we can express $m_3^0$ through $D$, the 
heaviest entry in $m_\nu^{\rm bimax}$ from Eq.\ (\ref{eq:mnubimax}). 
Then the above forms of the decay asymmetries can be rewritten as 
\bea 
\varepsilon_1^N \simeq 
\frac{\D -3 M_1}{\D 16 \pi \, v^2} \, |s| \, \sin 4 \varphi_6 ~,\\[0.3cm]
\varepsilon_1^\Delta \simeq  
\frac{\D -3 \, M_1}{\D 16 \pi \, v^2} \, |D| \, 
\sin 2 (2\beta + 2 \phi - \varphi_6 + \sigma)~.
\eea
As is should, the asymmetry $\varepsilon_1^N$ proportional to $m_\nu^{I}$ 
vanishes for $s = 0$, i.e., when there is no conventional 
type I see--saw term. 
It is seen that the contribution to the decay asymmetry stemming from the 
exchange of virtual Majorana neutrinos is suppressed in comparison  
to virtual triplet exchange by a typical factor of (ignoring phases) 
\be
\left|\frac{\varepsilon_1^N }{\varepsilon_1^\Delta} \right| 
\simeq  \left|\frac{s}{D} \right|~.
\ee
This is easily interpreted as the ratio of the the maximal 
entries in $m_\nu^{II}$ and $m_\nu^{I}$, respectively. 
This conclusion holds also when we use the full Dirac mass matrix 
Eq.\ (\ref{eq:mdCPV}). 
Since there are here (and in general) unmeasurable 
combinations of phases involved in $\varepsilon_1^\Delta$, 
however, it is possible that $\varepsilon_1^N$ dominates the 
baryon asymmetry though the main contribution to $m_\nu$ 
stems from $m_\nu^{II}$. 
In Refs.\ \cite{eap99,WRlep} a detailed bottom--up 
analysis of such scenarios can be found.  

\noindent Not surprisingly, without assuming any more 
simplifications of the mass matrices, the 
high energy $CP$ violation as required for leptogenesis in the 
decay asymmetries Eqs.\ (\ref{eq:epss}) decouples from the $CP$ violation 
as measurable at low energy in \meff{} or $J_{CP}$ 
as given in (\ref{eq:JCP1}). The same is true for the connection 
of low and high energy $CP$ violation in general type I see--saw 
models \cite{branco,PPR}. 

\noindent Note that within our framework 
the upper limit on the decay asymmetry is given by 
\be 
|\varepsilon_1^{\rm max}| \simeq 
\frac{3}{16 \pi} \, \frac{M_1}{v^2} (D + |s|) 
\simeq \frac{3}{16 \pi} \, \frac{M_1}{v^2} \, 
\left( \frac{m_3}{2} + \frac{5}{4} |s| \right)
\simeq 
\frac{3}{32 \pi} \, \frac{M_1}{v^2} \, \sqrt{\dma}~.
\ee
Hence, this upper limit is not much difficult (roughly a factor 2 weaker) 
from the usual bound in the conventional type I see--saw leptogenesis 
scenario \cite{DI}, 
$|\varepsilon_1^N| < \frac{3}{16 \pi} \, \frac{M_1}{v^2} \, \sqrt{\dma}$. 
This latter bound is valid for light neutrinos with normal 
or inverted mass hierarchy.   
In these cases, the limits on the decay asymmetry within the 
type I and type II see--saw mechanism are identical \cite{anki}. 
However, in the type II see--saw scenario, assuming 
quasi--degenerate light 
neutrinos with a common mass scale $m_0$, a bound of 
$|\varepsilon_1^{\Delta} + \varepsilon_1^N| 
< \frac{3}{16 \pi} \, \frac{M_1}{v^2} \, m_0$ 
can be derived \cite{anki}. 
This has to be contrasted with the limit 
$|\varepsilon_1^N| \ls \frac{3}{32 \pi} \, \frac{M_1}{v^2} \, 
\frac{\Delta m^2_{\rm atm}}{m_0}$ 
in case of leptogenesis within the type I see--saw 
mechanism \cite{DI}. Therefore, in case of type II see--saw the 
upper limit on the decay asymmetry for quasi--degenerate neutrinos is 
weaker by a factor of $2m_0^2/\dma$ with respect to the limit in 
case of the conventional type I see--saw. 
We remark that quasi--degenerate 
light neutrinos are more natural to obtain in type II see--saw scenarios.\\

\noindent 
One can check if the decay asymmetry has the 
correct order of magnitude to generate a sufficient baryon asymmetry. 
Let us assume for this purpose that the wash--out processes 
in our framework yield an efficiency factor for the 
lepton asymmetry of similar magnitude as in the usual conventional 
scenarios \cite{lepto03}. Solving the complete set of Boltzmann--equations 
is beyond the scope of this study. 
The overall scale of the decay asymmetries can be rewritten as 
\bea
\left|\varepsilon_1^N + \varepsilon_1^\Delta \right| \ls 
10^{-6} \left( \frac{\D M_1}{\D 10^{10} \, \rm GeV}\right) \, 
\left( \frac{\D m_3^0}{\D \sqrt{2 \cdot 10^{-3} \, \rm eV^2}} 
\right) \\[0.3cm] 
\sim 10^{-3} \left( \frac{\D v_R}{\D 10^{15} \, \rm GeV}\right)^2 \, 
\left( \frac{\D m_1^0}{\D 10^{-3} \, \rm eV} \right) \, 
\left( \frac{\D m_3^0}{\D \sqrt{2 \cdot 10^{-3} \, \rm eV^2}} \right)
~, 
\eea
where we used that $M_i = m_i \, v_R/v_L  $ and $v_R \, v_L \sim v^2$.  
The order of magnitude of the decay asymmetry typically 
required for succesful leptogenesis is about $10^{-6}$ \cite{lepto03}. 
Hence, as in the bottom--up analysis of \cite{eap99}  
in case of hierarchical light neutrinos and 
$m_D$ being related to the up--quarks, the decay asymmetry 
is for the ``natural'' value $v_R = 10^{15}$ GeV typically too large and 
requires some suppression by the phases or 
by somewhat smaller values of $v_R$. 

\section{\label{sec:concl}Summary}
The simple toy model presented in this paper serves to underline 
possible interplay of both terms in the type II see--saw formula. 
In type II see--saw scenarios it is possible that the 
conventional see--saw term $m_D^T M_R^{-1} m_D$ naturally gives only a 
small correction $s$ to the dominating triplet term $m_L$. It is 
tempting to assume that the dominating $m_L$ corresponds to 
bimaximal neutrino mixing. Then, as demonstrated in the present article, 
the small contribution $s$ from the conventional see--saw term can be 
sufficient to pull solar neutrino mixing away from being maximal. If this 
mechanism is realized, $|U_{e3}|$ and $1 - \sin^2 2 \theta_{23}$ 
receive corrections from zero of order 0.001 and 0.01, respectively. 
The presence of $CP$ violation does not change the typical 
behavior of those observables, except that atmospheric mixing can be 
allowed to be maximal. 
If the type I see--saw term is also 
responsible for generating the solar $\Delta m^2$, both 
$|U_{e3}|$ and $1 - \sin^2 2 \theta_{23}$ are significantly larger. 

\noindent The deviation from maximal solar neutrino mixing is 
described most conveniently and naturally via 
$U_{e2} = \sqrt{1/2}~(1 - \lambda)$ and the other deviations from bimaximal 
neutrino mixing will be proportional to appropriate powers of 
$\lambda$ as well. 
The ratio $R$ of the solar and atmospheric mass squared 
differences turns out to be of the order 
$\lambda^2$. This apparent coincidence is explained 
by the framework described in the present paper when one starts with 
vanishing \dms, because it holds that $\lambda \propto s$ and 
$R \propto s^2$. 

\noindent Since the type II term is consequence of a $SU(2)_L$ Higgs 
triplet term, this triplet can also contribute to the decay asymmetry in 
leptogenesis scenarios. The decay asymmetry produced by the 
exchange of virtual triplets is typically larger than the one 
produced by heavy Majorana exchange by a factor corresponding to 
the ratio of the maximal entries in $m_\nu^{II}$ and  $m_\nu^I$.

\vspace{0.5cm}
\begin{center}
{\bf Acknowledgments}
\end{center}
It is a pleasure to thank Sandhya Choubey for invaluable comments. 
This work was supported by the EC network HPRN-CT-2000-00152.

\newpage

\end{document}